\documentclass[journal]{IEEEtran}
\IEEEoverridecommandlockouts

\pdfoutput=1

\usepackage{amsmath, amssymb, amsfonts, amsthm}
\usepackage{algorithmic}
\usepackage{algorithm}
\usepackage{graphicx}
\usepackage{textcomp}
\usepackage{csquotes}
\usepackage{xcolor}
\usepackage{mhchem}
\usepackage{booktabs}
\usepackage{enumerate}
\usepackage[
backend=biber,
style=ieee,
citestyle=numeric-comp,
sorting=none,
dashed=false,
doi=true
]{biblatex}
\usepackage{hyperref}

\theoremstyle{remark}
\newtheorem*{remark}{Remark}

\def\BibTeX{{\rm B\kern-.05em{\sc i\kern-.025em b}\kern-.08em T\kern-.1667em\lower.7ex\hbox{E}\kern-.125emX}}

\begin{document}

\title{Accounting for Optimal Control in the Sizing of Isolated Hybrid Renewable Energy Systems Using Imitation Learning}

\author{Simon Halvdansson\textsuperscript{1}, Lucas Ferreira Bernardino\textsuperscript{1}, Brage Rugstad Knudsen\textsuperscript{1} \thanks{\textsuperscript{1}SINTEF Energy Research, Trondheim, Norway.\\

This publication has been produced with support from the LowEmission Research Centre (\url{www.lowemission.no}), performed under the Norwegian research program PETROSENTER. The authors acknowledge the industry partners in LowEmission for their contributions and the Research Council of Norway (296207).}}

\maketitle

\begin{abstract}
Decarbonization of isolated or off-grid energy systems through phase-in of large shares of intermittent solar or wind generation requires co-installation of energy storage or continued use of existing fossil dispatchable power sources to balance supply and demand. The effective \ce{CO2} emission reduction depends on the relative capacity of the energy storage and renewable sources, the stochasticity of the renewable generation, and the control of the isolated energy system. While the operation of the energy storage and dispatchable sources impacts the optimal sizing of the system, it is challenging to account for the effect of finite-horizon optimal control at the stage of system sizing. In this work, we present a flexible and computationally efficient sizing framework for energy storage and renewable capacity in isolated energy systems, accounting for uncertainty in the renewable generation and the optimal control. We implement an imitation learning approach to stochastic neural model predictive control (MPC) which allows us to relate the battery storage and wind peak capacities to the emissions reduction and investment costs while accounting for finite horizon, optimal control without solving an infeasible number of optimization problems. We evaluate the proposed sizing framework on a case study of an offshore energy system with a gas turbine, a wind farm and a battery energy storage system (BESS). In this case, we find a nonlinear, nontrivial relationship between the investment costs and the reduction in gas usage for different wind and BESS capacities.
\end{abstract}

\begin{IEEEkeywords}
Model predictive control, imitation learning, stochastic MPC, optimal sizing, isolated energy systems.
\end{IEEEkeywords}

\section{Introduction}
Isolated energy systems such as small islands, microgrids, offshore installations, remote communities, or research stations are characterized by their lack of a connection to central grids. In the absence of a base load supply, isolated energy systems often employ in situ fossil gas turbines or diesel generators \cite{Trevizan2021}. Increased attention is paid to decarbonizing isolated energy systems, both to reduce the \ce{CO2} emissions but also due to the often high fuel cost of diesel in remote areas. With the phasing-in of renewable intermittent power sources such as photovoltaic cells and wind turbines, sufficient balancing capacity is needed -- yet often obtained by combining energy storage and remaining balancing fossil-fueled backup sources \cite{Zhou2020a}. In this paper, we address the problem of cost-effective sizing of these hybrid fossil-renewable systems while accounting for the impact of optimal, feedback control of the uncertain renewable power sources during operation. This problem is nontrivial as it involves integrating finite-horizon stochastic control into a sizing problem of multiple technologies, possibly with a nonlinear cost function. However, the impact of the control policy including how weather forecasts are incorporated and how the uncertain generation is handled may impact system performance, thus motivating investigation of simplified, combined sizing and control approaches.   

Fixed costs or economies-of-scale effects in the investments together with generation uncertainty from the intermittent renewable energy sources cause the sizing of isolated, hybrid-renewable energy systems to be a stochastic, mixed-integer optimization problem. Common approaches to solve this class of problems include two-stage or multi-stage stochastic programming \cite{Alharbi2018} with Benders decomposition to reduce solution times \cite{Abdulgalil2019}, and bi-level optimization \cite{Rigo-mariani2017}. Other approaches include pattern search-based optimization methods with Monte Carlo simulations \cite{Arabali2014} and frequency-analysis approaches \cite{Makarov2012}. The first class of methods enables accounting for the uncertainty in renewable generation during the optimal system sizing. Yet, they normally assume perfect, possibly long-term foresight of the renewable generation and thereby fail to account for the effects of a finite prediction horizon, and feedback control, during the actual system operations. Frequency and sampling-based approaches account for the stochasticity and multiple scenarios of weather-dependent renewable generation, while neither account for the multi-variable control of the integrated renewable and energy storage system. Some contributions have been made to account for control in the design of microgrids, e.g. \cite{Zhou2020a}, however without rigorously addressing the stochastic control of the intermittent renewable sources. The small size and intermittent nature of the renewable power sources often available in isolated energy systems present new challenges in control and operations. This affects the planning stages of emissions reduction projects as assessing the impact of investment decisions on renewable power and energy storage requires accurate simulations of the system interactions and their optimal operations.

\begin{figure*}[!htbp]
    \centering
    \includegraphics[width=\linewidth]{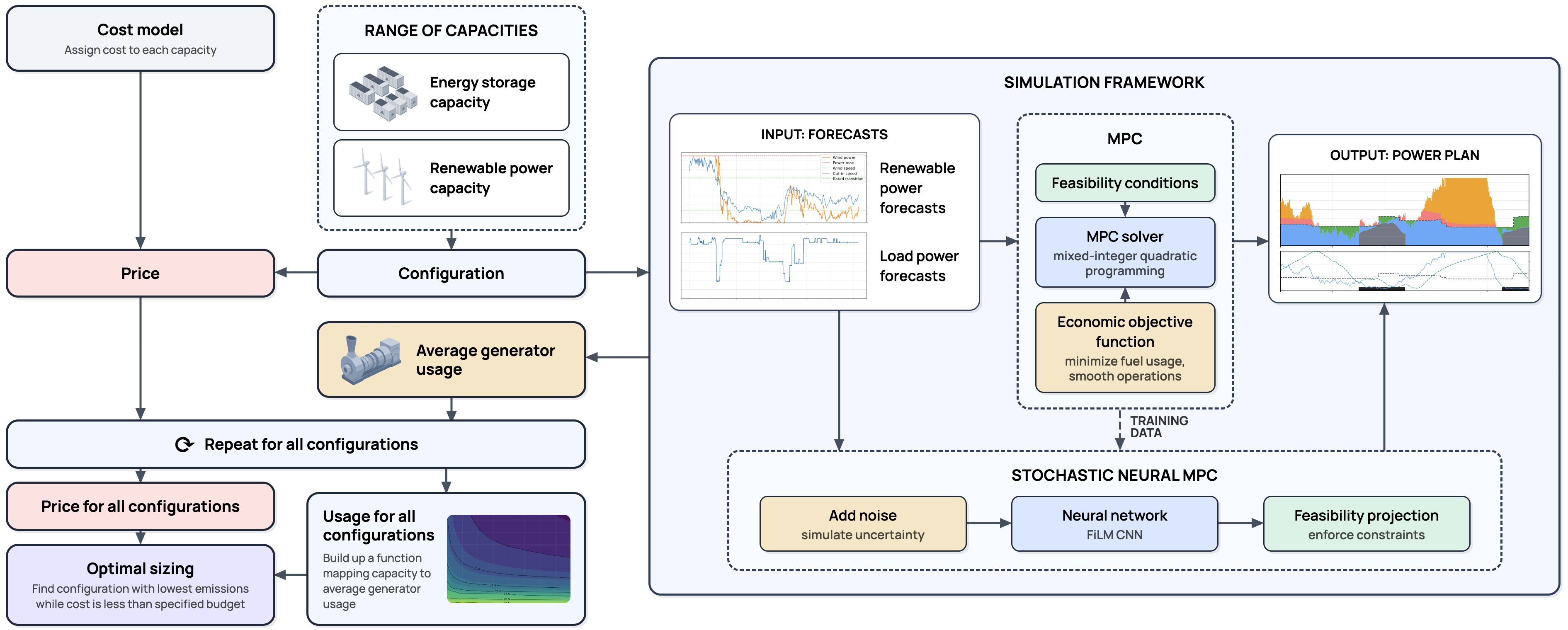}
    \caption{Flowchart overview of the control-aware sizing framework, starting with a cost model and the range of capacities considered and ending in the optimal sizing when generator usage and price for all configurations are known.}
    \label{fig:overview}
\end{figure*}

The increase in complexity is commonly dealt with by applying a more advanced control policy such as \emph{model predictive control} (MPC) in which we solve a mathematical dynamic optimization problem at each time step to determine the optimal next action. This has been done in the integration of different energy systems \cite{Turk2021}, control of an offshore energy system \cite{Hoang2024}, building energy control \cite{Gao2023}, and microgrid optimization \cite{Hu2024}, among others. Data generated by the control policy can then be used to inform the sizing of the associated energy system \cite{Su2023, AlQuraan2024}. We build on this line of work by considering the sizing of renewable capacity, not only battery capacity, and by using imitation learning on an MPC policy to improve runtime, enabling a simple grid search instead of a more involved search algorithm. This simplifies the process of finding the optimal configuration at a fixed investment budget. Neural MPC and MPC-guided imitation learning have recently been investigated for energy-system control \cite{Gao2022,Liu2025,Matrone2024,Irfan2026}. The methods in \cite{Gao2022,Liu2025,Matrone2024} use relatively simple multilayer perceptrons to imitate MPC policies, while \cite{Irfan2026} combines MPC-generated expert demonstrations with an LSTM policy for energy management in an islanded vehicle-to-grid microgrid. These studies primarily address operational control for prescribed system configurations and do not investigate control-aware joint sizing of renewable-generation and energy-storage capacities. Aiming to address these deficiencies, our contributions may be summarized as follows:
\begin{itemize}
    \item We develop a control-aware sizing framework for energy systems consisting of a power consumer, an intermittent stochastic renewable energy source, a dispatchable thermal generator which can provide backup power, and an energy storage system.
    \item We train a neural surrogate of the MPC policy over a range of storage and renewable capacities, enabling rapid estimation of residual fuel consumption across the design space. This allows energy-system designers to choose storage and renewable capacities using a lookup table under investment budget and emission reduction constraints, removing the need for an advanced search algorithm for optimal sizing and turning sizing from a CPU-bound problem to a GPU-bound one which scales significantly better.
\end{itemize}
A flowchart overview of the steps in determining the optimal sizing can be found in Figure \ref{fig:overview}.

In this way, we address the research question of how best to size an isolated hybrid renewable energy system which is modeled with a quadratic objective function. Our methodology combines an advanced control policy, MIQP MPC, with imitation learning to quickly search through the sizing configuration space. Our neural network is more expressive than those used in previous work, and can thus better represent the complex MPC policy. Feasibility is enforced through a robust projection step, and the resulting policy is incorporated into a stochastic planning method. These improvements make up the bulk of the novelty of the present work.

We demonstrate the proposed methodology on a case study on decarbonization of an offshore energy system. In that case, we substantially reduce the runtime compared to a standard MPC approach, allowing us to simulate a wide range of scenarios over different capacity combinations to develop a model for emissions reductions based on sizing, and provide investment advice under a prescribed cost model.

The remainder of this paper is organized as follows: In Section \ref{sec:background} we present the formulation of model predictive control and its extensions to imitation learning and taking into account uncertain forecasts. Section \ref{sec:framework} details our sizing framework, including our MPC setup, our neural network architecture, the algorithm for accounting for uncertainty and the investment input procedure. In Section \ref{sec:case_study} we apply this framework to a case study and in Section \ref{sec:conclusions} we offer concluding remarks.

\section{Background}\label{sec:background}
\subsection{Model predictive control}
At a high level, MPC controls a dynamical system to minimize an objective function, subject to constraints on the states the system is allowed to take. In MPC, we take a \emph{receding horizon} approach, meaning that we create a new control plan at each time step over a fixed-length time horizon and only implement the first action of the plan at each time step. In this way, we are able to ensure that we properly account for the long-term impact of our actions while simultaneously being able to rapidly respond to external inputs, system-model mismatch and new information entering our time horizon.

Formally, we split the system variables into three parts -- the state variables $x \in X$ which we can affect indirectly, the action variables $u \in U$ which we can choose freely and the exogenous variables $w \in W$ which are external. The system dynamics are modeled by a function $f : X \times U \times W \to X$ satisfying relation \eqref{eq:system_dynamics_general};
\begin{align}\label{eq:system_dynamics_general}
    x_{k+1} = f(x_k, u_k, w_k),
\end{align}
where a subscript $k$ corresponds to the value of the variables at the $k$-th time step. This allows us to recursively determine the state at any future time step $k+L$ given the actions at time $k, k+1, \dots, k+L-1$. Generally, each variable is a vector with either real or integer values.

For an MPC system we generally have constraints such as $X \subsetneq \mathbb{R}^N$ where $N$ is the number of variables in the state vector. In its most general form, we encode these constraints into two functions $g, h : X \times U \times W \to \mathbb{R}$ of the state, action and exogenous variables so that the constraints are satisfied if and only if $g \leq 0$ and $h=0$. The next critical component of an MPC system is the objective function $J$ which is defined over a time horizon of $H$ steps, i.e., $J : X^H \times U^H \times W^H \to \mathbb{R}$. For notational convenience, we will take a boldface letter to indicate a vector of the next $H$ time steps, i.e., $\boldsymbol{x}_k = (x_k, x_{k+1}, \dots, x_{k+H-1})$. The \emph{control policy} $\pi : X \times W^H \to U$ which determines the next action we should take is then defined implicitly by solving the minimization problem
\begin{equation}\label{eq:general_minimization}
    \begin{aligned}
        \min_{\boldsymbol{u}_k} \;& J(\boldsymbol{x}_k, \boldsymbol{u}_k, \boldsymbol{w}_k) \\
        \text{s.t.}\quad\;& x_{k+1} = f(x_k, u_k, w_k), \\
        & g(x_k, u_k, w_k) \leq 0, \\
        & h(x_k, u_k, w_k) = 0,
    \end{aligned}    
\end{equation}
and letting $\pi(x_k, \boldsymbol{w}_k)$ be $u_k$ from the minimizing argument, i.e., the first component of $\boldsymbol{u}_k$.

\subsection{Neural model predictive control}\label{sec:prelim_neural_mpc}
The minimization problem \eqref{eq:general_minimization} can be difficult to solve, especially when domains are not connected or one or more of $J$, $f$, $h$, or $g$ are nonlinear. One classical way of circumventing this issue is \emph{explicit MPC} \cite{Bemporad2002, Alessio2009} where the policy $\pi$ is formulated as a lookup table. Under some restrictive conditions on the problem \eqref{eq:general_minimization} (linear constraints, quadratic objective), it can be shown that the optimal control law is composed of polyhedra that partition the feasible domain. By precomputing these polyhedra, we essentially exchange computational cost for storage cost and can apply the optimal control policy with minimal inference latency. However, this solution scales poorly with the number of variables and is generally infeasible for larger systems.

A more modern approach to this problem is to train a neural network to approximate a known good control policy. This falls under the umbrella of \emph{neural MPC} \cite{Karg2020, Chen2018} and is called \emph{imitation learning} (IL). In IL, we learn to imitate an expert policy $\pi^*$ by formulating $\pi$ as a neural network and choosing its weights by minimizing $\Vert \pi^* - \pi \Vert$. Recent energy-management applications have similarly used MPC-generated state--action pairs to train recurrent imitation policies for microgrid dispatch \cite{Irfan2026}. However, without additional structure there are no guarantees that the output $\widehat{u} = \pi(x,w)$ will satisfy $g \leq 0$ and $h=0$. Solutions include adding regularizing terms to the objective function based on $g$ and $h$ \cite{Adhau2021} or manually changing the output $\widehat{u}$ by means of a projection $P : \widehat{u} \mapsto P(\widehat{u}) \in U$ to ensure that the constraints are satisfied \cite{Karg2020, Chen2018}.

\subsection{Stochastic model predictive control}\label{sec:stoch_mpc}
So far, we have treated the exogenous variables $w$ as fixed and determined ahead of time. When this is not the case, it is less clear how one should minimize the objective function $J$. This problem is the realm of \emph{stochastic} model predictive control (SMPC) \cite{Mesbah2016}. A common choice to deal with this uncertainty is to consider different scenarios in \emph{scenario-based} or \emph{multi-stage} MPC where we consider branching possibilities for how the system will develop in the future \cite{Lucia2013}. While popular, this approach suffers from an exponential complexity in the number of time steps to branch over. This can be counteracted by using the output of the scenario-based MPC as the training data for a neural MPC \cite{Lucia2018} but generating the training data will still suffer from the same increase in complexity and the resulting model will lack interpretability.

In this paper, we detail an alternative approach where we consider a collection of future scenarios, compute the optimal next action for each of them, rank the actions using a metric, and choose based on the quantile. This approach does not require evaluating the objective function in each scenario, meaning that a neural MPC policy can be used. Moreover, different scenarios can be evaluated in parallel, enabling GPU-acceleration. The policy is further detailed in Section \ref{sec:scenario_handling}.

\section{Control-aware sizing framework}\label{sec:framework}
In this section we detail our simulation framework in a general form to highlight its versatility and discuss how it can be applied to sizing. On a high level, we are interested in systems with a conventional power source, the use of which we aim to minimize, a renewable intermittent power source, an energy storage system and a power consumer. We will treat the consumer and the renewable power as external meaning we cannot affect them and we may not be able to predict them perfectly. Our framework is focused on controlling this system, both with a conventional control policy and a neural network, handling uncertainties in the input variables, and understanding the long-term average usage statistics of the system under different configurations to properly size the renewable components.

We first detail the setup and components of the simulation before outlining the sizing strategy. Finally we describe how the solver can be successively replaced by a stochastic neural MPC model.

\subsection{System setup and components}
Our controller uses a receding horizon approach where the current state and information about the next $H$ time steps are used to find optimal control actions at the current time step which are applied before moving to the next discrete time step. We let $T$ denote the total number of time steps in a complete scenario and write $k$ for the current time step.

Below we list the components of the system and their characteristic properties. Each component primarily plays a role through its power $P_k$.
\begin{itemize}
    \item \textbf{Load $\boldsymbol{P_k^\text{load}}$}
    In our energy system, the load comes from the power consumer which has deterministic values for the load at all points in time.
    \item \textbf{Dispatchable thermal generator $\boldsymbol{P_k^\text{DTG}}$}
    When renewable power is unavailable, we can use the dispatchable thermal generator to generate power. This system can be turned off and has a minimum power level when online.
    \item \textbf{Energy storage system $\boldsymbol{E_k^\text{ESS}}$, $\boldsymbol{P_k^\text{ESS}}$}
    To balance renewable power, we include an energy storage system. Charging and discharging have limited efficiency and the sign of the power will be set so that the power is positive when the ESS is charged and negative when discharged.
    \item \textbf{Renewable power source $\boldsymbol{P_k^\text{RPS}}$}
    The future power from the renewable power source will be based on historical data. In our final stochastic approach, we assume that we can sample various possible scenarios for this power source which is relevant in e.g., wind and solar setups where there exist forecasts for future wind speed and solar irradiance.
\end{itemize}

The primary objective of the control algorithm is to maintain a power balance where the load power can be met by a combination of the other three components. When the output power is greater than the required power, we need to curtail the excess generation. Formally, this is achieved via the introduction of an additional variable $P_k^\text{curt}$.

\subsection{Investment input}\label{sec:framework_investment}
Our motivation for the control scheme which we will set up is to inform investment decisions on the capacity of the renewable power source and the ESS. Specifically, the performance metric we wish to minimize is the average generator usage as a function of these two capacities
\begin{align}\label{eq:cost_g_func}
    G : \big(E_\text{max}^\text{ESS},\, P^\text{RPS}_\text{max}\big) \mapsto P^\text{DTG}_\text{avg}.
\end{align}
The exact interpretations of the capacities $E_\text{max}^\text{ESS}$ and $P^\text{RPS}_\text{max}$ are made clear in the next section. To estimate this average, we can consider a wide range of load and renewable scenarios and run an MPC for each one of them, averaging the results. Since each simulation runs over $T$ time steps and depends on the initial state-of-charge, we will choose $T$ to be large.

Investigating investment decisions requires a cost model. We can represent this as a fully general function $C$ of $E_\text{max}^\text{ESS}$ and $P^\text{RPS}_\text{max}$ which can include all of the expenses related to an investment, including the present value of future maintenance and appropriate penalties for using up limited space and weight budgets.

With these two functions in hand, we can make informed investment decisions. Provided that we can sample $G$ efficiently, it is straightforward to answer questions such as 
\begin{enumerate}[(i)]
    \item Given that the capacities are chosen optimally, what is the average DTG load for each cost?
    \item What are optimal energy storage and peak renewable capacities for given budget level or price point, provided that minimum average DTG load is the goal?
\end{enumerate}
These questions can be answered by running a grid search over $(E_\text{max}^\text{ESS},\,P^\text{RPS}_\text{max})$. We will demonstrate this for the case study in Section \ref{sec:case_study}.

\subsection{Control objective function and constraints}\label{sec:objective_and_constraints}
 With the convention that the energy storage system power is the derivative of its energy, we can write the total power balance equation as
\begin{align}\label{eq:power_balance}
    -P_k^\text{ESS} + P_k^\text{DTG} + P_k^\text{RPS} = P_k^\text{load} + P_k^\text{curt}.
\end{align}
Ensuring that the target load is always met is equivalent to only allowing positive curtailment, i.e.,
\begin{align}\label{eq:curt_positive}
    P_k^\text{curt} \geq 0.
\end{align}
Since the load needs to be met for all time steps, we set the limits for the controllable dispatchable thermal generator high enough that it alone can meet the target load. The fact that the generator is dispatchable, meaning it can be turned on and off and has a minimum power when online, is encoded as
\begin{align}\label{eq:dtg_power_limits}
    P_k^\text{DTG} \in \{ 0 \} \cup \big[P_\text{min}^\text{DTG}, P_\text{max}^\text{DTG}\big].
\end{align}
This can be accommodated by means of an auxiliary binary variable which encodes the on/off state at the cost of turning the optimization problem into a mixed-integer problem.

For the energy storage system we utilize an intermediate variable indicating the energy, $E_k^\text{ESS}$, which is subject to the limits
\begin{align}
    0 &\leq E_k^\text{ESS} \leq E_\text{max}^\text{ESS} \label{eq:ess_0100}
\end{align}
and the relation
\begin{align}
    E_k^\text{ESS} &= E_{k-1}^\text{ESS} + \Delta t \Big( \eta_\text{ch} P_{k,+}^\text{ESS} - \frac{1}{\eta_\text{dis}} P_{k, -}^\text{ESS} \Big) \label{eq:bess_charge}
\end{align}
where $\eta_\text{ch}$, $\eta_\text{dis} \leq 1$ are the charging and discharging efficiencies, and $P_{k,+}^\text{ESS},\, P_{k,-}^\text{ESS} \geq 0$ are the positive and negative parts of $P_k^\text{ESS} = P_{k,+}^\text{ESS} - P_{k,-}^\text{ESS}$, respectively. Note that for the $k=1$ relation, we need an initial condition $E_0^\text{ESS}$.

Lastly we place limits on the maximum charge/discharge rate of the ESS in the form
\begin{align}\label{eq:ess_power_limit}
    -P_\text{max}^\text{ESS} \leq P_k^\text{ESS} \leq P_\text{max}^\text{ESS}.
\end{align}
Equations \eqref{eq:power_balance}~-~\eqref{eq:ess_power_limit} are all the constraints we apply.

For our control objective function there are two terms which promote low and smooth usage of the dispatchable thermal generator;
\begin{align}\label{eq:loss_gt}
    \frac{\lambda_\text{usage}^\text{DTG}}{H}\sum_{k=1}^H P_k^\text{DTG} + \frac{\lambda_\text{smooth}^\text{DTG}}{H}\sum_{k=1}^H \big( P_k^\text{DTG} - P_{k-1}^\text{DTG} \big)^2
\end{align}
where we specify an initial value for $P_0^\text{DTG}$. Note the $1/H$ factor to make the same coefficients $\lambda$ work for different time horizons. To minimize degradation of the ESS, we penalize its usage quadratically. This means that the model is incentivized to charge/discharge the battery smoothly. We also reward the terminal battery level to encourage the model to treat the battery charge as a limited resource throughout the window as a form of value function. These incentives take the form
\begin{align}\label{eq:loss_bess}
    \frac{\lambda_\text{smooth}^\text{ESS}}{H} \sum_{k=1}^H \big( P_k^\text{ESS} \big)^2 - \lambda_\text{terminal}^\text{ESS} E_H^\text{ESS}.
\end{align}
Consequently, our objective function is of the form
\begin{equation}\label{eq:total_loss}
\begin{aligned}
    J &= \frac{\lambda_\text{usage}^\text{DTG}}{H}\sum_{k=1}^H P_k^\text{DTG} + \frac{\lambda_\text{smooth}^\text{DTG}}{H}\sum_{k=1}^H \big( P_k^\text{DTG} - P_{k-1}^\text{DTG} \big)^2\\
    &\hspace{4mm}+ \frac{\lambda_\text{smooth}^\text{ESS}}{H} \sum_{k=1}^H \big( P_k^\text{ESS} \big)^2 - \lambda_\text{terminal}^\text{ESS} E_H^\text{ESS}.
\end{aligned}
\end{equation}

In total, we can summarize our optimization task as follows: For each time step $k$, given $E_{k-1}^\text{ESS}$, $P_{k-1}^\text{DTG}$ and sequences $\boldsymbol{P}_k^\text{RPS}$ and $\boldsymbol{P}_k^\text{load}$, find sequences $\boldsymbol{P}_k^\text{DTG}$ and $\boldsymbol{P}_k^\text{ESS}$ such that \eqref{eq:power_balance}~-~\eqref{eq:ess_power_limit} hold, and which minimize \eqref{eq:total_loss}.

Since the dispatchable thermal generator on/off state needs to be encoded via a binary decision variable, this is a \emph{mixed-integer} optimization problem. Moreover, the objective function \eqref{eq:total_loss} is quadratic in $P_k^\text{ESS}$ and $P_k^\text{DTG}$ and all the constraints are linear. Hence the total problem is a \emph{mixed-integer quadratic programming} (MIQP) problem.

Note that at each time step, we have perfect knowledge with no uncertainty of the current wind which is what affects the power balance up until the next time step. As such, we will always be able to meet the power demand exactly. This simplification is made since we are more interested in the high-level behavior of our system, mostly with respect to the utilization of the dispatchable thermal generator. In a real installation, we would have another controller acting on shorter time scales to smooth out the short-term power variability.

Solving the MIQP problem iteratively with a receding horizon yields a valid MPC policy. While functional, the approach suffers from the problems mentioned in Section \ref{sec:background}. Before introducing uncertainties in the future load and renewable power, we first improve the runtime of the model by approximating it with a neural network in the next section.

\subsection{Imitation learning}\label{sec:imitation_learning}
The goal of our neural MPC model is to approximate the actual MPC policy detailed above while significantly reducing runtime, similar to the setup in \cite{Liu2025}. Since our ultimate goal is to use our control policy to compare the behaviors of our system with different ESS and renewable power capacities, we want to be able to condition the model on this. To enable this, we fix global maximum ESS and renewable power capacities and sample random capacities less than these during training. We condition the model on the fractions of the maxima during training and inference. To deal with possibly different output ranges based on these scales, we will restrict the model to output in the $[0,1]$ range for the dispatchable energy source and the $[-1, 1]$ range for ESS power and then rescale by $P_\text{max}^\text{DTG}$ and $P_\text{max}^\text{ESS}$ respectively.

\subsubsection{Training data}\label{sec:training_data}
Having defined the optimal control problem, we generate training data by evaluating the MPC policy at randomly sampled operating conditions. Specifically, we learn the mapping
\begin{align}\label{eq:next_step_mapping}
    \big( E_{k-1}^\text{ESS},\, P_{k-1}^\text{DTG},\,
    \boldsymbol{P}_k^\text{RPS},\, \boldsymbol{P}_k^\text{load} \big)
    \mapsto
    \big( P_k^\text{ESS},\, P_k^\text{DTG} \big),
\end{align}
conditioned on the ESS and renewable capacity scales.

For each training sample, the ESS and renewable capacity scales are drawn independently from a uniform distribution on $[0,1]$. Conditional on the resulting ESS capacity, the initial ESS energy is drawn uniformly between zero and its maximum capacity. The DTG is initially off with probability $0.5$; otherwise, its previous power is drawn uniformly between its minimum and maximum operating powers.

The load and renewable forecasts are generated by independently selecting uniformly random starting points in their respective historical data sets and extracting contiguous windows of length $H$. The renewable data are then converted to power and scaled by the sampled renewable capacity. For each sampled input to \eqref{eq:next_step_mapping}, we solve the horizon-$H$ MIQP and retain only the first optimal action. A new operating condition is sampled for every solve, producing a diverse expert data set that can be generated in parallel.

\subsubsection{Model architecture}\label{sec:model_arch}
\begin{figure*}[!htbp]
    \centering
    \includegraphics[width=0.89\linewidth]{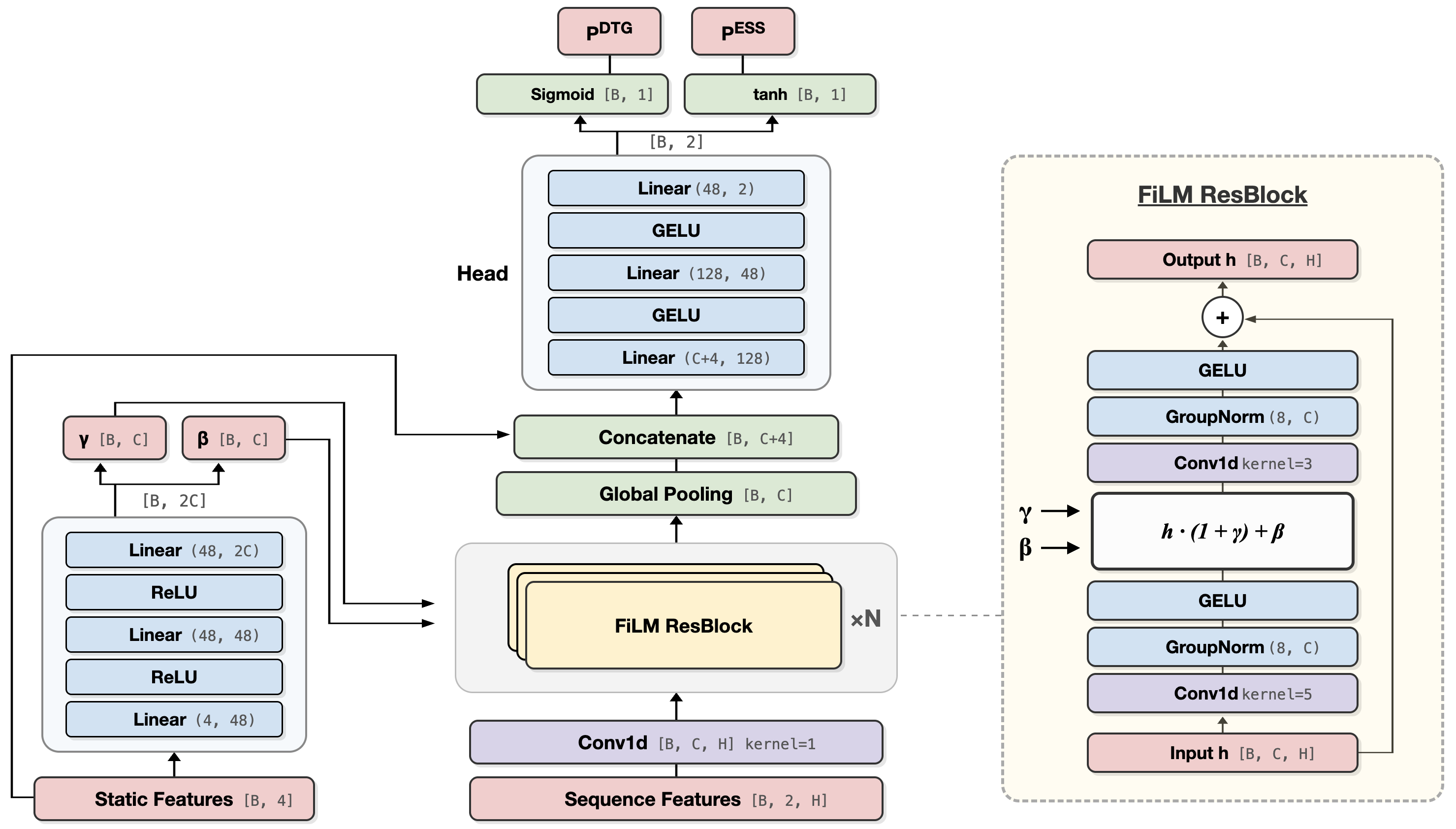}
    \caption{Detailed view of the architecture of the neural network used in the neural MPC, starting with the sequence features (future load and renewable power) and static features (current state, capacity). The parameters $B, C, H$ are the batch size, number of hidden channels and sequence length respectively.}
    \label{fig:arch}
\end{figure*}

To learn the control problem, we train a ResNet-v2-style \cite{he2016identity} convolutional neural network (CNN) with GroupNorm, GELU activations, and feature-wise linear modulation (FiLM) \cite{perez2018film} to encode the current ESS state-of-charge and DTG usage, as well as the ESS and renewable capacities. The architecture is illustrated in detail in Figure \ref{fig:arch}. The future load and renewable power are treated as a 2-channel time series which is first mapped to a tensor of shape $(C,H)$ where $C$ is the number of hidden channels, by means of a convolutional layer with kernel size $1$. The FiLM conditioning takes the form of two vectors of length $C$; $\gamma,\, \beta$, which are the output of a simple 2-layer multilayer perceptron (MLP). They are applied inside each ResNet block to modify the latent representation $h$ as $h \mapsto h(1+\gamma) + \beta$. After the final block, we take the mean over $H$ to get a vector of length $C$ which we connect to an output head which is a 2-layer MLP which outputs two numbers. We finally apply a hyperbolic tangent to one of the outputs and a sigmoid mapping to the other and these are $\widehat{P_k^\text{ESS}}$ and $\widehat{P_k^\text{DTG}}$ respectively. The hyperbolic tangent and sigmoid allow us to properly account for their power limits \eqref{eq:ess_power_limit} and \eqref{eq:dtg_power_limits}.

\subsubsection{Action projection}\label{sec:action_projection}
As discussed in Section \ref{sec:prelim_neural_mpc}, since the neural network output does not exactly match the deterministic MPC algorithm, there is no guarantee that the actions it recommends are valid in the sense of conforming to the constraints \eqref{eq:power_balance}~-~\eqref{eq:ess_power_limit}. To solve this problem, we apply simple post-processing of the outputs $\widehat{P_k^\text{DTG}},\, \widehat{P_k^\text{ESS}}$ to ensure that the constraints hold. We refer to this processing as a projection since it maps to the space of valid actions and is the identity on that space.

The most important constraint is the power balance equation \eqref{eq:power_balance}. Since negative curtailment is not allowed, in the case that
\begin{align}\label{eq:negative_curtailment}
    P_k^\text{RPS} + \widehat{P_k^\text{DTG}} - \widehat{P_k^\text{ESS}} < P_k^\text{load},
\end{align}
either $\widehat{P_k^\text{DTG}}$ needs to be increased or $\widehat{P_k^\text{ESS}}$ needs to be decreased, while still adhering to \eqref{eq:power_balance}~-~\eqref{eq:ess_power_limit}, turning \eqref{eq:negative_curtailment} into an equality. This can be achieved by dispatching power from the ESS with contingencies for using the thermal generator or both depending on the power and energy limits. Similarly, to ensure that \eqref{eq:dtg_power_limits} holds, we map network outputs for the dispatchable thermal generator power in $\big[0, P_\text{min}^\text{DTG}/2\big)$ to $0$ and outputs in $\big[P_\text{min}^\text{DTG}/2, P_\text{min}^\text{DTG}\big)$ to $P_\text{min}^\text{DTG}$.

On top of ensuring that all the constraints hold, we can further increase the performance of the neural MPC solution by explicitly forbidding simultaneous battery discharge and curtailment since this will never be the case in any solution of \eqref{eq:general_minimization}. We view these corrective actions as the final step of the neural MPC so that we have a guarantee that the outputs of the neural MPC algorithm are always valid actions.

\subsection{Scenario handling of uncertainties}\label{sec:scenario_handling}
With the neural MPC model set up, we move towards integrating it into a custom stochastic MPC algorithm. Specifically, given that the neural MPC inference is quick and only returns the next optimal action, we propose the following stochastic MPC algorithm based on an economic robustness metric $r : U \to \mathbb{R}$ and a parameter $\alpha \in [0,1]$:
\begin{enumerate}
    \item Sample $N$ different future scenarios via their corresponding exogenous variables $\{ \boldsymbol{w}_k^n \}_{n=1}^N$.
    \item Compute the next action recommended by $\pi$ for each scenario as
    \begin{align}\label{eq:action_collection}
        \mathcal{A}_N = \big\{ \pi(x_k, \boldsymbol{w}_k^n) \big\}_{n=1}^N.
    \end{align}
    \item Sort the actions $\mathcal{A}_N$ based on the economic robustness metric $r$.
    \item Choose an action at the $\alpha$-quantile in the sorted list.
\end{enumerate}
\begin{remark}
    Step 2, computing \eqref{eq:action_collection}, can be done in parallel, and if $\pi$ is a neural MPC policy we can stack the scenarios along the batch dimension, increasing performance substantially.
\end{remark}

Note in particular that the economic robustness metric is a function of a single action. Since we are never at risk of not being able to meet the power demand, our notion of economic robustness is related to minimizing the risk of running out of stored energy and having to ramp up the DTG rapidly due to unexpected future low renewable production. To hedge against this scenario, we should run the DTG in the near term and charge the ESS. As such, we choose 
\begin{align}\label{eq:sort_alpha_loss}
    r(P^\text{DTG},\, P^\text{ESS}) = P^\text{DTG} + P^\text{ESS}.
\end{align}
One interpretation of this formula is that it is the ``safe" power which we activate to be in a better position in the future.

As an example, consider the case where there is a $10\,\text{MW}$ power deficiency after accounting for the renewable power which needs to be met by the dispatchable thermal generator and/or discharging the ESS. In a risk-averse strategy, we would not want to discharge the ESS and instead run the DTG. Both of these choices contribute to increasing the value of $r$. Similarly, in the case where renewable power is greater than the load target, we would most likely not run the DTG at all and instead charge the ESS. If we are risk-averse, we might opt to charge the ESS over a shorter time horizon, since we are unsure if the strong winds will stay, meaning that $P_k^\text{ESS}$ will be higher which corresponds to a higher value of $r$ again.

Under the assumption that the proposed scenarios accurately reflect the future, this algorithm allows us to properly account for our risk appetite via the quantile parameter $\alpha$ where a higher value of $\alpha$ means that we adopt a policy which is more economically robust. While this is a general and easy-to-implement approach, it does require the auxiliary function $r$. However, the case can be made that defining the risk through an explicit economic robustness metric rather than implicitly through the objective function is preferable since the objective function is formulated to describe the optimal action in a single scenario rather than relative notions such as ``take the action which is more risk-averse than needed in 90\% of plausible scenarios".

\begin{remark}
    Related ideas for stochastic MPC can be found in the literature under names including \emph{multi-forecast} MPC \cite{Shen2021_mfmpc} and \emph{multi-scenario} MPC \cite{Tian2019_msmpc}. However, to the best of our knowledge, the setup as described here of computing entirely separate optimal next actions for a collection of scenarios and then choosing one based on some ranking has not previously been investigated. In particular, we have not found the amenability of neural MPC algorithms with batching to this problem discussed elsewhere. 
\end{remark}
\subsection{Algorithmic summary}
In Algorithm \ref{alg:neural_mpc} we have collected the main steps of setting up and running the neural MPC, with the exception of the neural network training as this is a standardized process.

\begin{algorithm}[ht]
\small
\caption{Training and deployment of the stochastic neural MPC}
\label{alg:neural_mpc}
\begin{algorithmic}[1]
\REQUIRE Historical data $\mathcal{D}$, horizon $H$, expert samples
$M$, scenarios $N$, and quantile $\alpha$
\ENSURE Trained neural policy $\pi_\theta$

\STATE Initialize expert dataset $\mathcal{D}_{\mathrm{exp}}$

\FOR{$i=1,\ldots,M$}
    \STATE Sample capacities
    $c_i=\left(P_{\max}^{\mathrm{RPS}},E_{\max}^{\mathrm{ESS}}\right)$
    \STATE Sample state
    $x_k=\left(E_{k-1}^{\mathrm{ESS}},P_{k-1}^{\mathrm{DTG}}\right)$
    \STATE Sample forecasts
    $w_k=\left(\boldsymbol{P}_{k}^{\mathrm{load}},
    \boldsymbol{P}_{k}^{\mathrm{RPS}}\right)$
    \STATE Solve the MIQP in
    \eqref{eq:power_balance}--\eqref{eq:total_loss}
    \STATE Extract the first optimal action
    $u_k^*=\big(P_k^{\mathrm{ESS},*},P_k^{\mathrm{DTG},*}\big)$
    \STATE Add $(x_k,w_k,c_i,u_k^*)$ to $\mathcal{D}_{\mathrm{exp}}$
\ENDFOR

\STATE Normalize and split $\mathcal{D}_{\mathrm{exp}}$
\STATE Train $\pi_\theta$ using Huber loss

\FOR{each simulation step $k$}
    \STATE Sample scenarios
    $\{\boldsymbol{P}_{k}^{\mathrm{RPS},n}\}_{n=1}^{N}$
    \STATE Evaluate $\pi_\theta$ for all scenarios in parallel
    \STATE Rescale and project the predicted actions
    \STATE Rank the actions using
    $r(u)=P^{\mathrm{DTG}}+P^{\mathrm{ESS}}$
    \STATE Apply the action at quantile $\alpha$
    \STATE Update the system state
\ENDFOR
\end{algorithmic}
\end{algorithm}

\section{Case study: Decarbonizing an offshore energy system}\label{sec:case_study}
In this case study we consider a hypothetical offshore energy system on the Norwegian continental shelf (NCS), where the goal is to decarbonize the energy generation. The system is powered by wind turbines, with a single gas turbine for backup, and a battery energy storage system (BESS).

\subsection{System overview}\label{sec:case_offshore_system_overview}
We first give an overview of the power producers and consumers on the platform. 
\begin{itemize}
    \item \textbf{Load source $\boldsymbol{P_k^\text{load}}$} We have access to proprietary energy usage data from a platform on the NCS provided by one of the operators in the LowEmission centre \cite{LowEmission}. This data is sampled at $\Delta t = 1\,\text{h}$ but we can resample it to align with the $\Delta t$ of the renewable power source. A subset of the load data is displayed in Figure \ref{fig:load_month}. As can be seen in the figure, the nominal power consumption is around $30\,\text{MW}$.
    
    \begin{figure}[!htbp]
        \centering
        \includegraphics[width=0.90\linewidth]{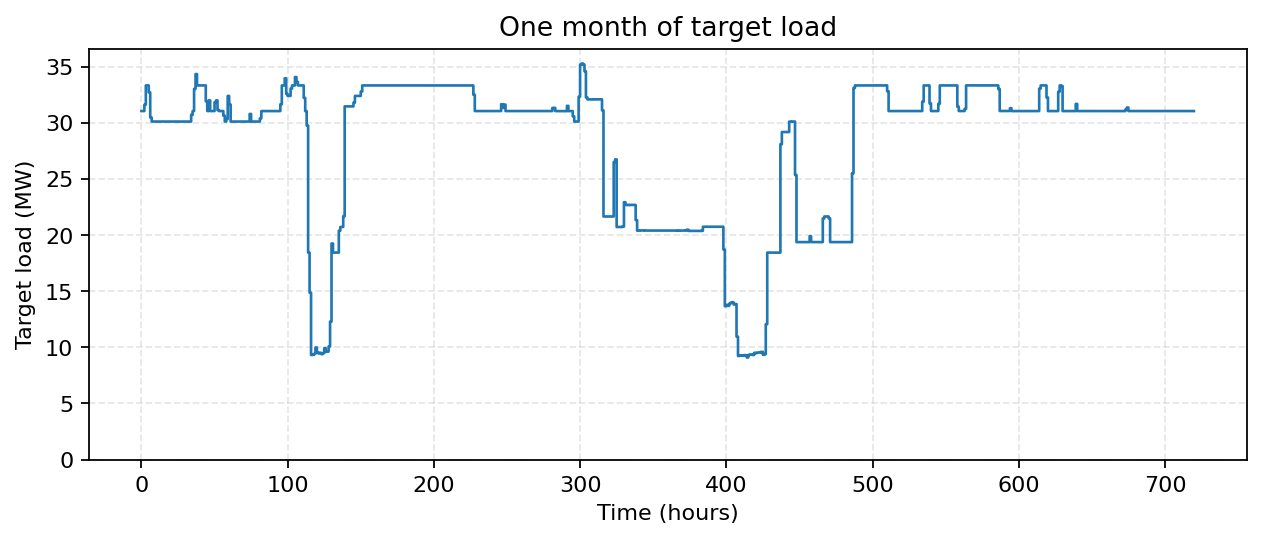}
        \caption{Example of the target load data over a month.}
        \label{fig:load_month}
    \end{figure}
    
    \item \textbf{Dispatchable thermal generator $\boldsymbol{P_k^\text{GT}}$} The dispatchable thermal generator for our system is a single gas turbine. We set $P_\text{max}^\text{GT} = 40\,\text{MW}$ and $P_\text{min}^\text{GT} = 0.35\cdot 40\,\text{MW}$ to approximately match a conventional offshore gas turbine \cite{Siemens2023SGT750}. Rapid ramping of the gas turbine is discouraged by the smoothness term in the objective function.
    
    \item \textbf{Energy storage system $\boldsymbol{E_k^\text{BESS}}$, $\boldsymbol{P_k^\text{BESS}}$} A realistic form of energy storage system for an offshore energy system platform is a battery energy storage system (BESS). We set $\eta_\text{ch} = \eta_\text{dis} = 0.96$ for roundtrip efficiency of approximately $92\%$ which is consistent with industry numbers for Li-ion batteries \cite{Kebede2022}. We choose a $C$-rate of $1$ hour which means that $P_\text{max}^\text{BESS} = \frac{E_\text{max}^\text{BESS}}{1\,\text{h}}$.
    
    \item \textbf{Renewable power source $\boldsymbol{P_k^\text{WTG}}$}
    We use a wind turbine generator (WTG) as our renewable power source and its interaction with the complete system is purely as an exogenous signal. As a data source we use wind speed observations from 2020 at the Brage offshore platform \cite{MET_wind_data}. The wind data can then be converted to power via the cubic relation $P_k^\text{WTG}(v) = A_3 v^3 + A_2 v^2 + A_1 v + A_0$  with constants from \cite{IEA15MW_ORWT} for wind speeds above a set cut-in speed, as a constant $P_\text{max}^\text{WTG}$ above the rated wind speed, and as $0$ below the cut-in or above the cut-out speed. A subset of the wind speed data and the associated wind power data is shown in Figure \ref{fig:wind_72}.
\begin{figure}[!htbp]
    \centering
    \includegraphics[width=0.95\linewidth]{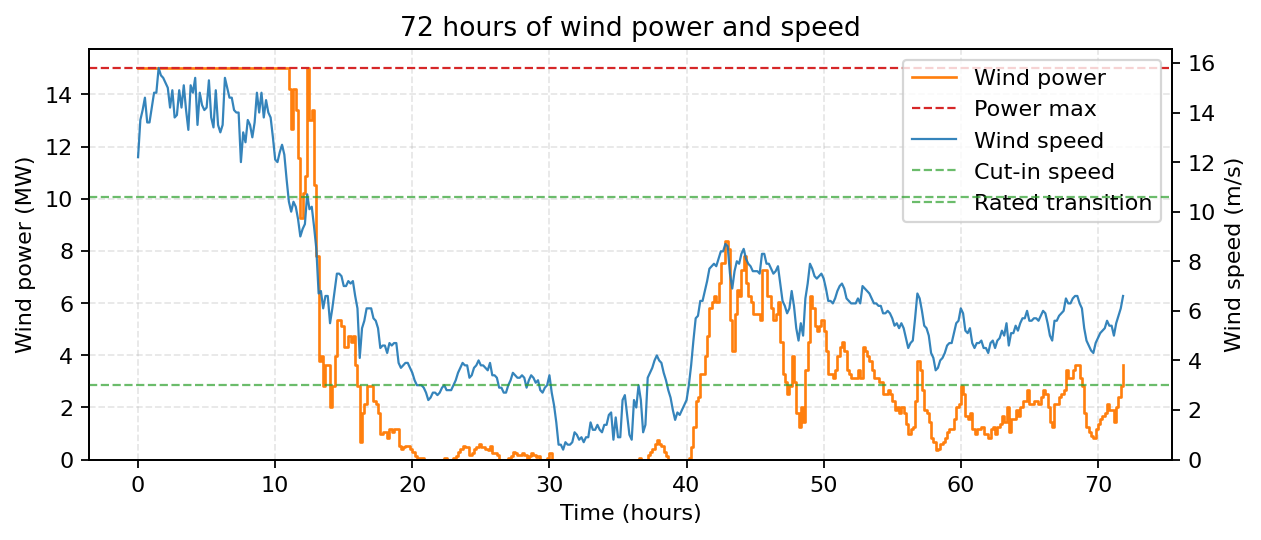}
    \caption{Example of wind speed and power data over a three-day period.}
    \label{fig:wind_72}
\end{figure}
The sampling interval of the data is $\Delta t = 10\,\text{min}$ so we use this for all of our time series.
\end{itemize}

\begin{figure*}[!htbp]
    \centering
    \includegraphics[width=0.72\linewidth]{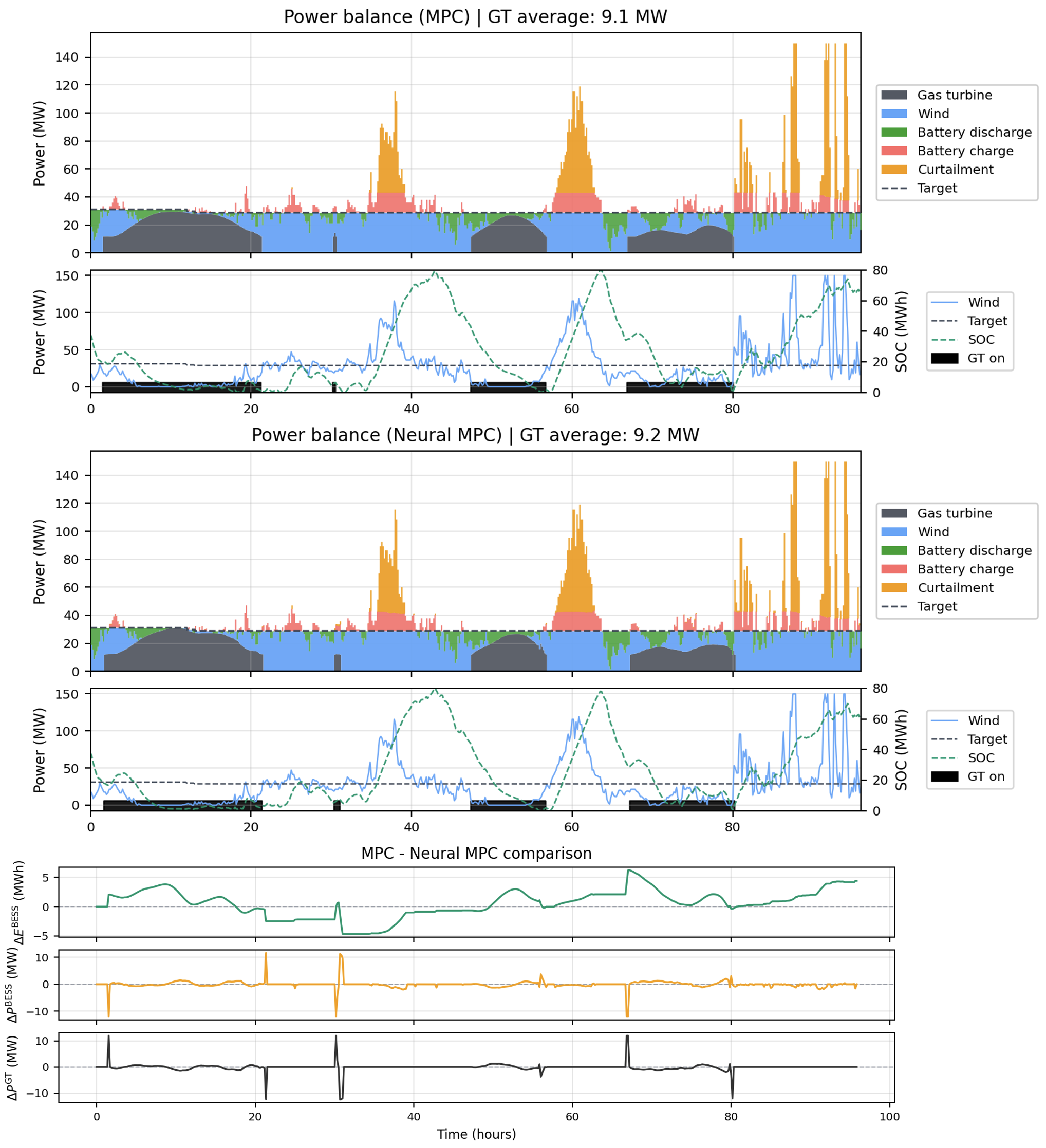}
    \caption{Power plans for standard MPC (top) and neural MPC (middle) for the same scenario. The bottom subfigure highlights the discrepancies in BESS energy, power and gas turbine power between standard and neural MPC.}
    \label{fig:mpc_vs_nmpc_comp}
\end{figure*}

\subsection{MPC results}
In this section we discuss the MPC algorithm applied to the energy system of the case study. 

\subsubsection{Deterministic MPC}
Through manual tuning, we have found the parameter values in \eqref{eq:constants} to be appropriate for the deterministic MPC algorithm but remark that the exact values here are of little consequence.
\begin{equation}\label{eq:constants}
\begin{aligned}
    \lambda_\text{usage}^\text{GT}&= 1, &\lambda_\text{smooth}^\text{BESS} &= 2\cdot 10^{-3},\\
    \lambda_\text{smooth}^\text{GT} &= 0.6,&\lambda_\text{terminal}^\text{BESS} &= 10^{-2}.
\end{aligned}
\end{equation}
We have also chosen a look-ahead window of $6$ hours which corresponds to $H =  6\,\text{h} / \Delta t = 36$ time steps. For the examples in this section we set $P_\text{max}^\text{WTG} = 150\,\text{MW}$ and $E_\text{max}^\text{BESS} = 80\,\text{MWh}$ which is large enough that we see large amounts of interaction between the different system components.

 The MIQP problem is solved using the \emph{Solving Constraint Integer Programs} (SCIP) solver \cite{SCIP}, controlled using Python through PySCIPOpt \cite{pysciopt}. In the top subfigure in Figure \ref{fig:mpc_vs_nmpc_comp} we see the result of the MPC control policy for a selected scenario. In that figure, we see the effect of the objective function in the smooth ramping of the gas turbine, the discharging of the battery around hour $45$ to delay starting the gas turbine, and the even charging rate leading to full charge during periods of high wind.

\subsubsection{Neural MPC}
To train our neural MPC algorithm, we sample $600\,000$ actions from the MPC algorithm specified above with input data chosen randomly as specified in Section \ref{sec:training_data}. For the model described in Section \ref{sec:model_arch}, we choose a hidden channel dimension of $96$, $4$ blocks, hidden conditioning dimension $48$, and a head with hidden dimensions $(128, 48)$, resulting in a model with a total of $327,954$ parameters. These values were selected using a hyperparameter search performed with Optuna \cite{akiba2019optuna}. The model is implemented in PyTorch \cite{paszke2019pytorch}.

We train with Huber loss for $50$ epochs with the AdamW optimizer, a batch size of $512$ and a learning rate of $4\cdot 10^{-4}$. This leads to a validation Huber loss of $0.14$ and a mean absolute error (MAE) of $0.29\,\text{MW}$. Note however that due to the discontinuous nature of gas turbine usage, we should expect the errors to be extremely heavy-tailed which is why we choose Huber loss instead of mean squared error. See the bottom part of Figure \ref{fig:mpc_vs_nmpc_comp} for an example of the errors in a full rollout. Pure MLP, LSTM and transformer-based methods were also considered and found to perform worse than the conditioned CNN.

The neural MPC model generally performs very similarly to the deterministic MPC, especially after the action projection described in Section \ref{sec:action_projection}. This can be seen by comparing the two action sequences in Figure \ref{fig:mpc_vs_nmpc_comp}. In the bottom subfigure, we see the exact differences which are generally concentrated to the exact time step during which the gas turbine is started and stopped. Note also that the average gas turbine usage ($9.1\,\text{MW}$ and $9.2\,\text{MW}$) are very close; this is the metric which is relevant for sizing decisions.

In Table \ref{tab:mpc-runtime} we compare the runtime of the neural and standard MPC algorithm for a four-day example consisting of 576 time steps. It should be remarked that the standard MPC has large variability in its runtime due to the use of an MIQP solver. Meanwhile the neural MPC algorithm performs the same computation regardless of the input parameters and is consequently much more stable in its runtime. In the table we see that the neural MPC algorithm offers a substantial speedup which can be increased further by running on a GPU. We remark also that compression techniques such as quantization \cite{jacob2018quantization} and pruning \cite{han2015learning} could likely reduce the runtime of this algorithm further without a significant loss in accuracy.

\begin{table}[ht]
    \centering
    \caption{Runtime comparison for $T=576$ time steps over a four-day period.}
    \label{tab:mpc-runtime}
    \begin{tabular}{lll}
        \toprule
        Algorithm & Hardware & Runtime \\
        \midrule
        Standard MPC & MacBook Pro M2       & $120$ s \\
        Neural MPC   & MacBook Pro M2 (MPS) & $4$ s \\
        Neural MPC   & NVIDIA RTX 5070 Ti   & $1.4$ s \\
        \bottomrule
    \end{tabular}
\end{table}

\subsubsection{Stochastic neural MPC}\label{sec:stochastic_nmpc}
We model uncertainty in the renewable generation via a collection of scenarios for the future. To generate these scenarios, we take the historical wind speed for the next $H$ time steps, add a random walk with standard deviation $0.5\,\text{m/s}$ and clip the resulting wind speed at zero. We then compute wind power using the cubic relation described in Section \ref{sec:case_offshore_system_overview}. This means that if the wind speed is far below the wind turbine cut-in, different scenarios may not change the wind power, which happens frequently in practice. 

\begin{figure}[!htbp]
    \centering
    \includegraphics[width=0.72\linewidth]{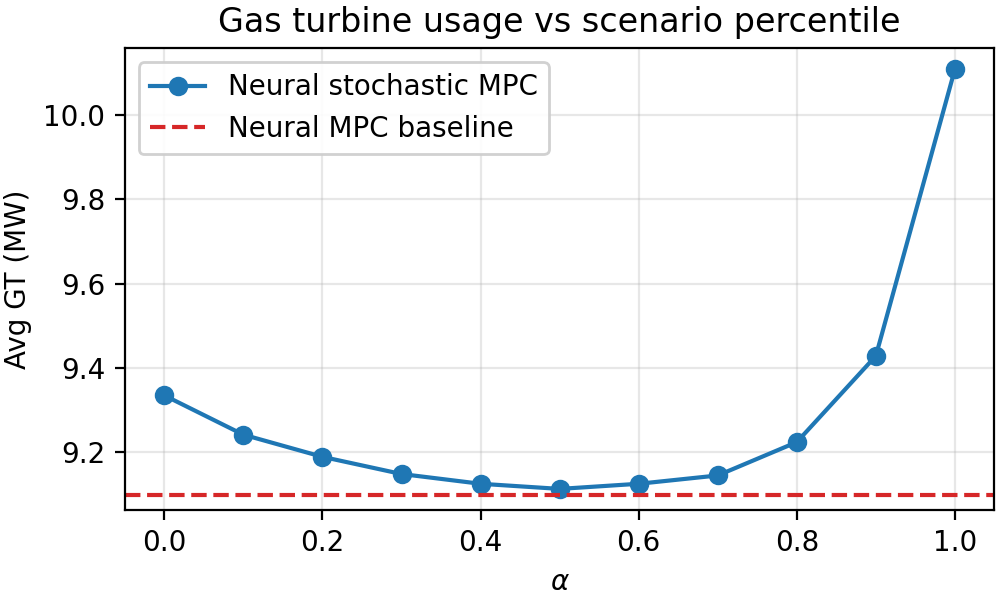}
    \caption{Gas turbine usage as a function of the risk parameter $\alpha$. The dashed red line indicates the neural MPC with exact forecast as input.}
    \label{fig:alpha_dependency}
\end{figure}

In implementing the algorithm described in Section \ref{sec:scenario_handling}, we choose $\alpha = 0.6$ for a slightly risk-averse strategy and $N=64$ scenarios. The resulting average gas turbine utilization over a collection of scenarios for different values of $\alpha$ is illustrated in Figure \ref{fig:alpha_dependency}. Here we see that choosing a high quantile, meaning being more pessimistic about future wind power, generally leads to higher gas turbine usage. Similarly, being overly optimistic also leads to increased gas turbine usage. The choice $\alpha=0.5$, corresponding to the median action with respect to our robustness metric, leads to the lowest average gas turbine usage with this specific scenario-generation procedure. 

\begin{figure*}[!htbp]
    \centering
    \includegraphics[width=0.7\linewidth]{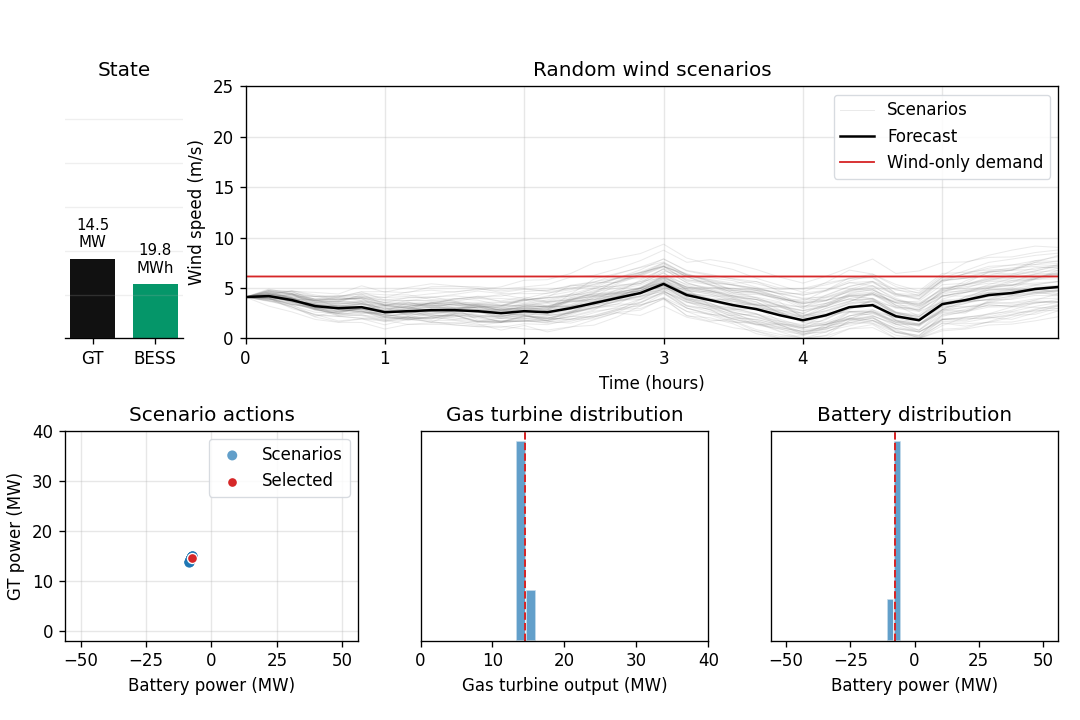}
    \caption{Snapshot of a single time step for the neural stochastic MPC process and its view of the time horizon. In the top row is the current state and the collection of possible wind scenarios based on the forecast, together with the red line indicating the wind speed required to power the platform completely by wind. In the bottom row is a scatter plot as well as histograms of the suggested gas turbine and battery powers for the different scenarios.}
    \label{fig:nsmpc_snapshot}
\end{figure*}

In Figure \ref{fig:nsmpc_snapshot}, the considered wind scenarios and the associated actions as given by the neural MPC algorithm are visualized as histograms and a scatter plot with the actions in the $\alpha$-quantile selected with respect to \eqref{eq:sort_alpha_loss} highlighted in red.

In the top half of the figure, we see all the considered wind scenarios as well as the historical data they are based on, all starting from the current wind speed. It is the top row, i.e., the current gas turbine usage, battery state-of-charge and future wind profile, as well as the future target load which is the input to the neural network. 

In Table \ref{tab:control-hyperparameters} we have collected all relevant parameters for the case study and framework.

\begin{table}[ht]
    \centering
    \caption{Selected configuration parameters for the MPC and imitation-learning algorithms.}
    \label{tab:control-hyperparameters}
    \begin{tabular}{lll}
        \toprule
        Component & Parameter & Value \\
        \midrule
        \multicolumn{3}{l}{\textit{Deterministic MPC}} \\
        & Sampling interval, $\Delta t$ & $10\,\mathrm{min}$ \\
        & Prediction horizon, $H$ & $36$ steps ($6\,\mathrm{h}$) \\
        & GT usage weight, $\lambda_\mathrm{usage}^\mathrm{GT}$ & $1$ \\
        & GT smoothing weight, $\lambda_\mathrm{smooth}^\mathrm{GT}$ & $0.6$ \\
        & BESS smoothing weight, $\lambda_\mathrm{smooth}^\mathrm{BESS}$ & $2\times10^{-3}$ \\
        & Terminal BESS weight, $\lambda_\mathrm{terminal}^\mathrm{BESS}$ & $10^{-2}$ \\
        & Optimization solver & SCIP \\
        & Solver interface & PySCIPOpt \\
        \addlinespace
        \multicolumn{3}{l}{\textit{Imitation learning}} \\
        & Number of expert samples & $600{,}000$ \\
        & Network architecture & ResNet-v2 CNN \\
        & Hidden channels & $96$ \\
        & Residual blocks & $4$ \\
        & Conditioning dimension & $48$ \\
        & Output-head dimensions & $(128,48)$ \\
        & Trainable parameters & $327{,}954$ \\
        & Loss function & Huber loss \\
        & Optimizer & AdamW \\
        & Learning rate & $4\times10^{-4}$ \\
        & Batch size & $512$ \\
        & Training epochs & $50$ \\
        \addlinespace
        \multicolumn{3}{l}{\textit{Stochastic neural MPC}} \\
        & Number of scenarios, $N$ & $64$ \\
        & Risk quantile, $\alpha$ & $0.6$ \\
        \bottomrule
    \end{tabular}
\end{table}

\subsection{Sizing results}
\begin{figure*}[!htbp]
    \centering
    \includegraphics[width=0.72\linewidth]{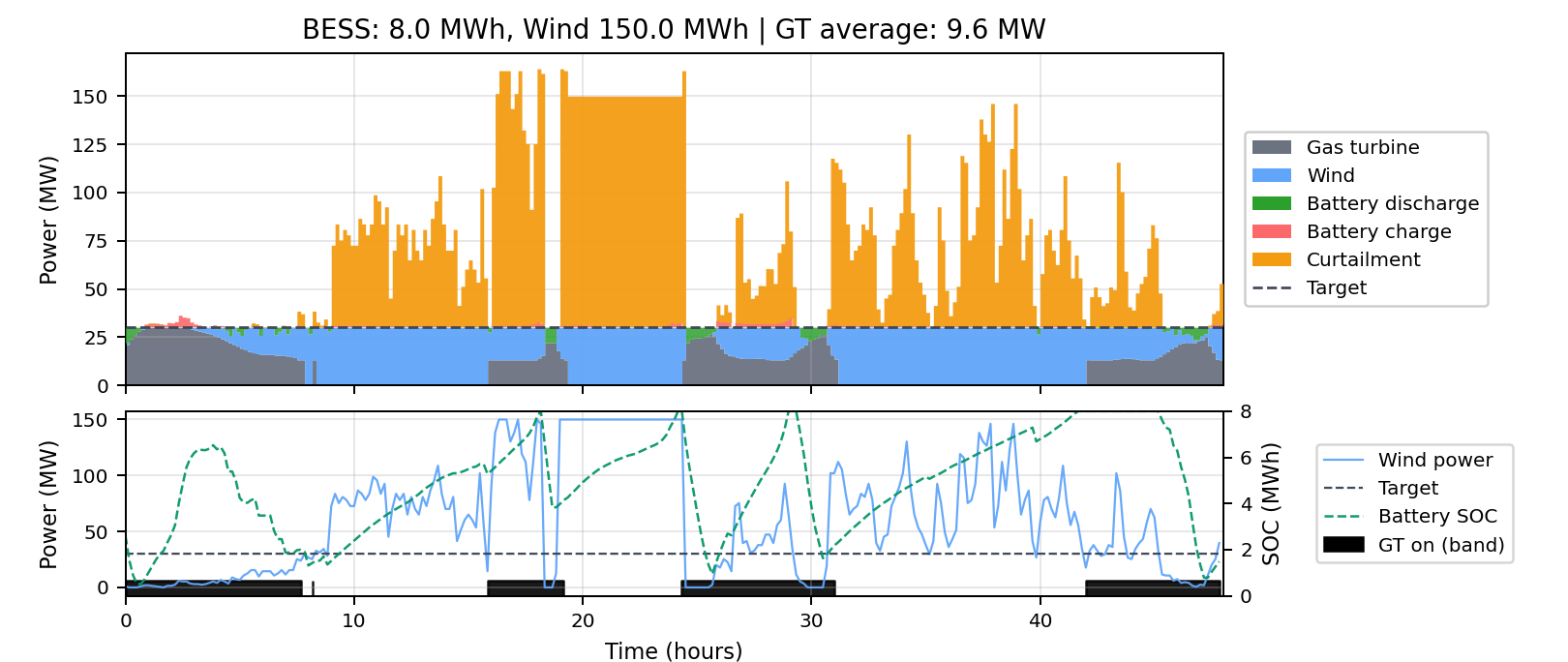}
    \includegraphics[width=0.72\linewidth]{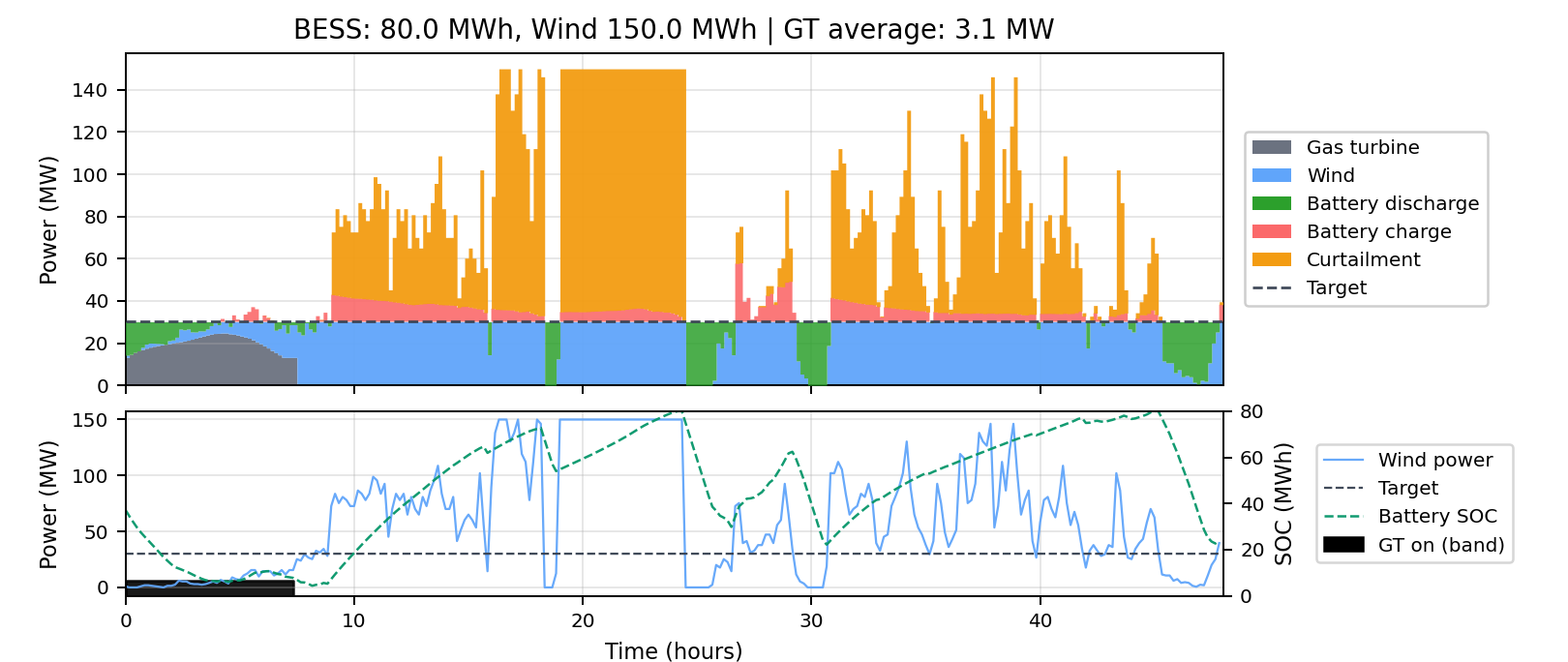}
    \caption{The same wind and load scenario simulated with an $8\,\text{MWh}$ (top) and an $80\,\text{MWh}$ (bottom) battery.}
    \label{fig:capacity_comparison}
\end{figure*}

Recall that we trained all our algorithms to handle differently sized BESS units and wind farms. In Figure \ref{fig:capacity_comparison}, we compare the same scenario with BESS capacities of $8\,\text{MWh}$ and $80\,\text{MWh}$. There we see that with the larger battery we are able to avoid turning on the gas turbine in three out of the four time periods with low wind and thereby decreasing the average gas turbine usage from $9.6\,\text{MW}$ to $3.1\,\text{MW}$.

\begin{figure*}[!htbp]
    \centering
    \includegraphics[width=0.82\linewidth]{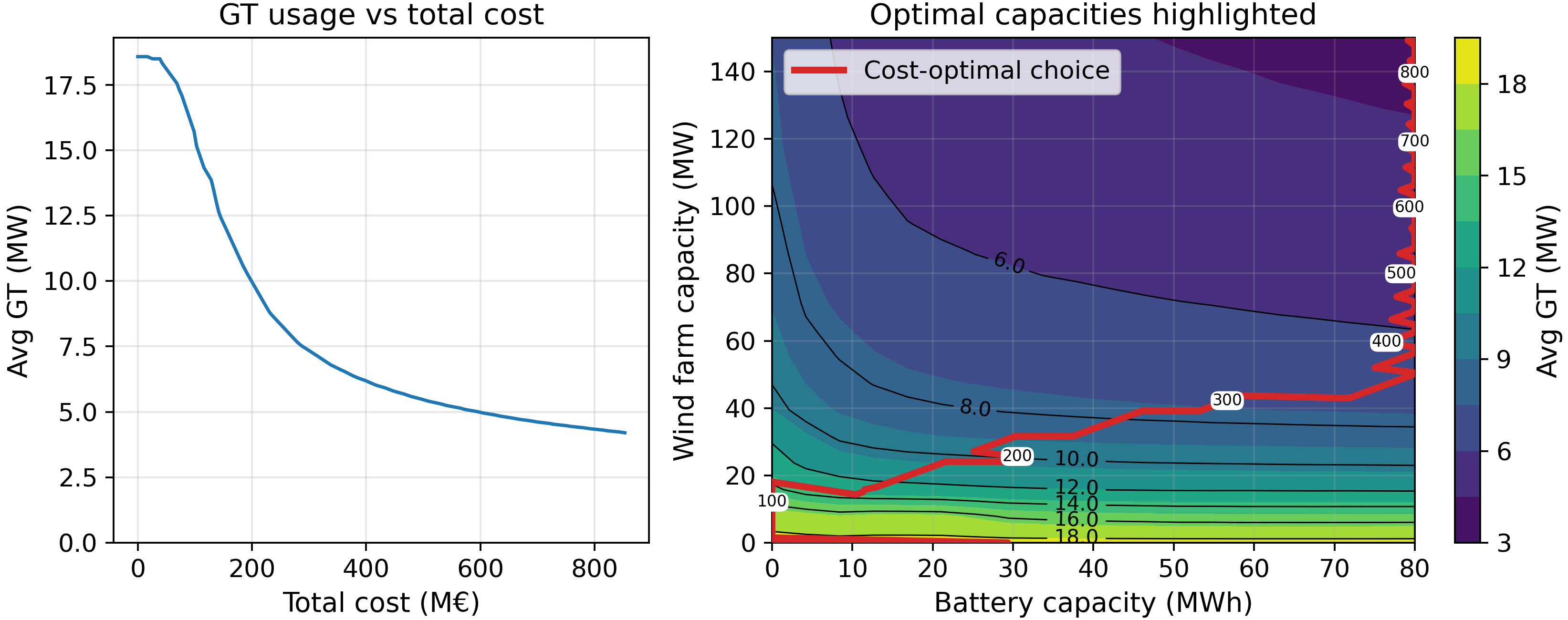}
    \caption{Optimal investment and results for a given cost model for wind and BESS capacity. Left: Average gas turbine usage as a function of total investment in wind and BESS with cost-optimal choices. Right: Contour plot of average gas turbine usage as a function of battery and wind capacity. Overlaid in red is the optimal investments at different price points, with multiples of $100\,\text{M€}$ highlighted.}
    \label{fig:optimal_per_price}
\end{figure*}

For the investment costs, we use an affine cost function with marginal and fixed costs given by
\begin{equation}\label{eq:cost_function}
\begin{aligned}
    C(E^\text{BESS}_\text{max}, P^\text{WTG}_\text{max}) &= A_\text{BESS}E^\text{BESS}_\text{max} + B_\text{BESS}\\
    &+ A_\text{WTG}P^\text{WTG}_\text{max} + B_\text{WTG}.
\end{aligned}
\end{equation}
Using rough estimates based on publicly available data \cite{NREL2024_CostOfWind, BNEF2024_BESScost} while accounting for additional costs due to offshore deployment, we set $A_\text{WTG} = 5\,\text{M€}/\text{MW}$, $B_\text{WTG} = 35\,\text{M€}$ and $A_\text{BESS} = 0.6\,\text{M€}/\text{MWh}$, $B_\text{BESS} = 20\,\text{M€}$. We remark that these values and the affine structure of \eqref{eq:cost_function} are illustrative and may not capture the complete, location-dependent costs of offshore installations. Applying the proposed imitation learning methodology in Section \ref{sec:imitation_learning}, running extensive simulations like those in Figure \ref{fig:capacity_comparison} for different wind and BESS capacities, we estimate the average gas turbine usage \eqref{eq:cost_g_func} and relative investment costs as shown in the left of Figure \ref{fig:optimal_per_price}. For each combination of these wind and BESS capacities, we find the corresponding investment costs and average gas turbine usage illustrated to the right in Figure \ref{fig:optimal_per_price}. Note that when either capacity is zero we skip the constant cost component.

In the left of Figure \ref{fig:optimal_per_price} we see that the very lowest investments have limited effects on the emissions reductions which is due to the fixed cost of wind power. Specifically, the model can initially only utilize the BESS which can be seen from the horizontal red line in the bottom left corner of the right subfigure. Once the budget is enough to accommodate the fixed cost of wind power, we see a rapid decline in emissions up to around $250\,\text{M€}$ after which the effect tapers off, indicating that this could be an efficient investment level. In the right subfigure, the labels indicate costs in millions of euros, starting at the bottom left corner. In this particular example, the optimal investment path is to add wind turbines up to the nominal platform load and then scale up batteries together with wind capacity. Note that BESS capacity becomes saturated at the $80\,\text{MWh}$ limit around $350\,\text{M€}$.

In a real scenario, one would most likely use a non-affine cost model to account for the limited weight and space on an offshore platform. Still, the high-level insights from this analysis are likely to remain relevant. We have seen that a mixture of wind and batteries can be cost competitive with a purely wind-based approach at the cost assumptions considered here.

\section{Concluding remarks}\label{sec:conclusions}
We set up a framework for the optimal sizing of isolated energy systems which uses accurate estimates of the fuel consumption for a wide class of systems by learning an MPC policy and running it for a collection of scenarios. This framework was then applied in a case study enabling decision makers to identify optimal investment capacities for wind and BESS given a budget level. A direct output of the framework is the expected emission reduction, by means of resulting average gas turbine usage, accounting for a large set of weather scenarios and stochastic MPC system operations.

The implemented sizing methodology applies an exhaustive exploration of the decision space to obtain an explicit map of expected generator usage and investment costs. The problem described in Section \ref{sec:framework_investment} can be formally treated as a multi-objective optimization problem, to which for instance direct search algorithms can be applied for systems with more technologies to be evaluated for the investments. We chose an exhaustive search to highlight the significant speedups enabled by moving to a GPU-accelerated MPC algorithm which was approximately $80$ times faster for a single scenario and parallelizes significantly better than a classical CPU-based MPC algorithm for the scenario handling of uncertainties. The switch to a GPU-based algorithm for exploring the decision space also enables new methods for speeding up inference additionally such as quantization and batching of different configurations in the same forward pass.

The main limitation of the present work is in showing empirically that the approach generalizes well to other situations. While neural MPC performed well in the case study discussed, we have not implemented the method for a wide class of cases or cases where the underlying MPC problem is of higher complexity. Our method is dependent on the availability of ample training data which can be more expensive for MPC problems of higher complexity. Another possible weakness is our scenario handling of uncertainties; there are different approaches to stochastic MPC which we have not implemented in our framework. Future work could focus on implementing alternative stochastic MPC algorithms such as scenario-based multi-stage MPC discussed in Section \ref{sec:stoch_mpc} and considering different case studies, including standard benchmark MPC problems for a more detailed performance comparison.

The imitation learning method for the stochastic neural MPC model discussed in the paper has the potential to be utilized for analysis of other systems and control problems, while the sizing framework is applicable for many types of isolated energy systems including systems with both PV and wind energy, mixed storage technologies, and diesel instead of gas turbines for backup. Our architecture and setup for imitation learning neural MPC are formulated in a general way and are likely to be useful for future work in the field.

\section*{Acknowledgments}
The authors thank Saket Adhau for insightful discussions during the early stages of this work.

\printbibliography

\end{document}